\documentclass[aps,prx,twocolumn,groupedaddress,showpacs,floatfix,superscriptaddress,longbibliography]{revtex4-2}
\usepackage[plainpages=false,pdfpagelabels,colorlinks=true,linkcolor=red,urlcolor=blue,citecolor=blue,pdftitle={Title},pdfauthor={},pdfdisplaydoctitle=true,pdfduplex=DuplexFlipLongEdge]{hyperref}
\usepackage{epsfig}
\usepackage{amsmath,amssymb}
\usepackage{physics}
\usepackage{bm}
\usepackage{graphicx}
\usepackage[dvipsnames,usenames]{color}
\usepackage[normalem]{ulem}
\usepackage{soul}
\usepackage{notes2bib}
\usepackage{xcolor}

\bibnotesetup{
note-name = ,
use-sort-key = false
}

\newcommand\numberthis{\stepcounter{equation}\tag{\theequation}}

\begin{document}

\title{Quantum State Transfer in Interacting, Multiple-Excitation Systems}
\author{Alexander Yue} 
\affiliation{Department of Physics and Astronomy, University of California, Davis, CA 95616, USA}
\author{Rubem Mondaini} 
\affiliation{Department of Physics, University of Houston, Houston, Texas 77004, USA}
\author{Qiujiang Guo}
\affiliation{School of Physics, ZJU-Hangzhou Global Scientific and Technological Innovation Center, and Zhejiang Key Laboratory of Micro-nano Quantum Chips and Quantum Control, \\ Zhejiang University, Hangzhou, China}
\author{Richard T. Scalettar}
\affiliation{Department of Physics and Astronomy, University of California, Davis, CA 95616, USA}

\date{\today}


\begin{abstract}

Quantum state transfer (QST) describes the coherent passage of quantum information from one node in a network to another. Experiments on QST span a diverse set of platforms and currently report transport across up to tens of nodes in times of several hundred nanoseconds with fidelities that can approach 90\% or more. Theoretical studies examine both the lossless time evolution associated with a given (Hermitian) lattice Hamiltonian and methods based on the master equation that allows for losses. In this paper, we describe Monte Carlo techniques which enable the {\it discovery} of a Hamiltonian that gives high-fidelity QST.  We benchmark our approach in geometries appropriate to coupled optical cavity-emitter arrays and discuss connections to condensed matter Hamiltonians of localized orbitals coupled to conduction bands. The resulting Jaynes-Cummings-Hubbard and periodic Anderson models can, in principle, be engineered in appropriate hardware to give efficient QST.

\end{abstract}

\maketitle


\section{Introduction}\label{sec:Intro}

Given a Hamiltonian $\hat {\cal H}$, the time evolution of a quantum mechanical system from an initial wave function $|\psi (0) \rangle$ is governed by the unitary map $|\psi(t) \rangle = \hat {\cal U}(t)  |\psi(0) \rangle = e^{-{\rm i} \hat {\cal H} t} |\psi(0) \rangle$ \cite{hbar}. One expects, generically, that wave functions spread with time since their initial state encodes uncertainty in momentum as well as position.  The most simple example is that of a free particle in one dimension where, if $|\psi (0) \rangle$ is spatially localized with uncertainty $(\Delta x)(0)^2 = \sigma_x^2$, the subsequent development will be such that  $(\Delta x)(t)^2 = \sigma_x^2 \big[\, 1 + \hbar t/ 2 m \sigma_x^2 \, \big]$. At long times $t$ the position uncertainty is proportional to $\sqrt{t}$, reflecting the analogy between the  diffusion and the (imaginary time) Schr\"odinger equations. As a consequence, achieving ``high-fidelity Quantum State Transfer,'' (QST)~\cite{cirac97} in which an excitation is transported from one particular spatial location to another {\it without spreading}, requires special engineering.  Solving this problem for appropriate hardware is essential to a host of potential applications in quantum computing ~\cite{divincenzo00,beals13,preskill18}.

At the experimental level, QST is being explored in a variety of realizations: from atom-photon coupling in optical cavities \cite{stute13}, Yb atoms connected by optical fibers \cite{olmschenk09},  optical cavities connected by waveguides \cite{vogell17},  a chain of superconducting qubits with tunable nearest-neighbor couplings \cite{li18}, and the implementation of {\it phonon}-mediated QST \cite{bienfait19}. Theoretical studies of QST encompass a similar goal -- to determine a Hamiltonian $\hat {\cal H}$ that evades the usual wave packet spreading. This program inverts the more usual quantum mechanical problem of solving for the time evolution operator $\hat {\cal U}$ of a {\it given} $\hat {\cal H}$, to the determination of an appropriate $\hat {\cal H}$ which leads to a desired $\hat {\cal U}$.

One illustration of this process is the prescription of Christandl {\it et al.}~\cite{christandl04} who showed that a one-dimensional hopping Hamiltonian with link-dependent hybridization $J_i = J_0 \sqrt{i\,(N-i)}$ exhibits perfect QST of a state initially localized at site $i=1$ to arrival with unit probability to site $i=N$ at a time $t=\pi/(2 J_0)$. Here $\{\, J_i\, \}$ for $i=1, 2, \ldots ,(N-1)$ are the hoppings from site $i$ to site $i+1$ and $N$ is the chain length. Many existing schemes for QST \cite{kay2006perfect,kay2010,keele2022combating}, like that of Christandl, focus on single excitations (or multiple, non-interacting excitations) in one dimension. However, experiments have begun to probe many-body effects in QST \cite{kandel21, Diepen2021}, as well as higher dimensions~\cite{QST_our_exp}. Naturally, the theoretical determination of a Hamiltonian that incorporates correlations between several particles is a challenging task.  In this paper, we describe a general Monte Carlo method to compute Hamiltonians which result in high-fidelity QST.

Our paper is outlined as follows: Section \ref{sec:Methods} describes the class of model Hamiltonians that define the space over which our Monte Carlo searches and the method itself. Here we emphasize an analogy between coupled cavity arrays and the periodic Anderson model (PAM) of condensed matter physics. Section \ref{sec:Single-Excitation} then provides some initial illustrations of how our approach works, demonstrating that it can reproduce solutions to the single excitation systems discussed in Ref.~\cite{christandl04}. It also benchmarks how the number of Monte Carlo steps scales with target fidelity and system chain length. 

Section \ref{sec:Multiple Excitations} turns to the main focus of the paper -- the unsolved problem of achieving high-fidelity QST with multiple interacting excitations. Here, in the Jaynes-Cummings-Hubbard (JCH) model, the mixed bosonic and fermionic statistics of photons and emitters, respectively, induce many-body effects. In the case of the PAM, it is the on-site repulsion that introduces correlations, and one must constrain the Monte Carlo search to prevent the simulation from finding trivial, non-interacting solutions. As we advance, Sec.~\ref{sec:Multiple Transfers} considers what happens when we try to achieve high fidelity QST between multiple pairs of initial and final states 
    \bibnote{The Christandl prescription~\cite{christandl04} automatically accomplishes perfect QST between all pairs representing transitions between states located symmetrically about the chain center, as might be inferred from the symmetry of the $\{ \, J_i \, \}$.}. 
Lastly, Sec.~\ref{sec:Conclusions} summarizes our results and describes the outlook for related calculations.


\section{Models and Methodology}\label{sec:Methods}

In this section, we will describe the systems we consider and define their Hamiltonians. We will also characterize the computational methods we use to tune coupling rates to achieve high-fidelity QST.


\subsection{Models}

The Jaynes-Cummings-Hubbard model~\cite{Angelakis2007} describes the coupling of photons and atom-like emitters in a lattice of optical cavities. Photons can move between nearest-neighbor cavities, and within each cavity, they can be absorbed or emitted by atoms in that cavity. In this paper, we focus on one-dimensional lattices and on systems in which every cavity contains exactly one emitter. With these constraints, the JCH Hamiltonian is 

\begin{align*} \label{eq:JCHH}
\hat {\cal H}_{\textnormal{JCH}} &= \sum\limits_{i=1}^{N}
\Omega_i^{\phantom{\dagger}} \hat a_{i}^{\dagger} \hat a_{i}^{\phantom{\dagger}} + 
\sum\limits_{i=1}^{N-1} J_{i}^{\phantom{\dagger}} \big( \, \hat a_{i+1}^{\dagger} \hat a_{i}^{\phantom{\dagger}}
 + \hat a_{i}^{\dagger} \hat a_{i+1}^{\phantom{\dagger}} \, \big) \\
&+\sum\limits_{i=1}^{N} \omega_{i}^{\phantom{\dagger}} \hat \sigma_{i}^{+} \hat \sigma_{i}^{-}
 + g_{i} \big( \, \hat a_{i}^{\dagger}  \hat \sigma_{i}^{-}  + \hat \sigma_{i}^{+} \hat a_{i}^{\phantom{\dagger}} \, \big) \ .\numberthis
\end{align*}

\begin{figure}
\hskip-0.1in
\includegraphics[width=0.7\columnwidth]{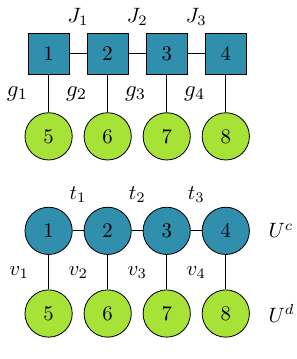}
\caption{
    \underbar{Top:}
    Geometry of the one-dimensional Jaynes-Cummings-Hubbard Model with $N=4$ cavities each containing one emitter. Photons can ``hop'' between adjacent cavities (squares) $i$ and $i+1$ via the coupling rates $J_{i}$. Within cavity $i$ a photon can be absorbed and excite emitter $i+N$ (circles) via the coupling rates $g_{i}$. The same process can be reversed, traveling from emitter to cavity via $g_{i}$. The numbers indicate our convention for labeling the states in the Hilbert space, i.e.~the rows of our matrix in the single excitation sector.
    \underbar{Bottom:}
    Geometry of the Periodic Anderson Model with $N=4$ sites. Spin up and spin down electrons can ``hop'' between adjacent conduction orbitals $i$ and $i+1$ via the hybridizations $t_{i}$ and between a conduction orbital $i$ and localized orbital $i+N$ via the hybridizations $v_{i}$, but not between adjacent localized orbitals. When two electrons occupy the same conduction or localized orbital, they have an on-site repulsion, $U^c, U^d$, respectively. 
}
\label{fig:JCHGeom} 
\end{figure}

We define the length of the system, $N$, as the number of cavities. $\hat a_{i}^{\dagger}\big(\hat a_{i}^{\phantom{\dagger}} \big)$ are photon creation (annihilation) operators in cavity $i$ and  $\hat \sigma_{i}^{+}\big(\hat \sigma_{i}^{-} \big)$ are excitation (de-excitation) operators for the emitter in cavity $i$. The model is parameterized by cavity energies $\Omega_i$, inter-cavity photon hopping rates $J_{i}$, emitter energy levels $\omega_{i}$ and photon-emitter coupling rates $g_{i}$. The geometry of such a system can be seen in Figure \ref{fig:JCHGeom} (top). Initial studies of time evolution in the JCH model in the single-excitation sector have recently been presented in \cite{PhysRevB.105.195429}.  We will consider both situations in which the couplings are constrained to be symmetric about the chain center and ones in which they vary freely. It is important to note that we treat $J_i, g_i$ as experimentally tunable parameters, but assume that $\Omega_i, \omega_i$ remain fixed. As might be expected, this choice can be important depending on whether the states between which the QST is occurring
obey the same symmetry.

To solve the JCH model~\eqref{eq:JCHH}, and the associated time evolution operator $\hat {\cal U}$, one can employ exact diagonalization. Since $\hat {\cal H}_{\rm JCH}$ conserves the total number of excitations, $N_{\rm exc} \equiv \sum_{i} \langle\big( \hat a_{i}^{\dagger} \hat a_{i}^{\phantom{\dagger}} + \hat \sigma_{i}^{+} \hat \sigma_{i}^{-}\big)\rangle$, the matrix for $\hat {\cal H}_{\rm JCH}$ is block diagonal, and we can consider each excitation number subspace individually. For the $N=4$ geometry illustrated in Fig.~\ref{fig:JCHGeom}(top), the Hilbert space dimension is ${\cal D}=8$ when $N_{\rm exc}=1$, since the single excitation can be a photon in any of the cavities or an excited atom. ${\cal D} = 32$ when $N_{\rm exc}=2$ since there are four states with two photons in a single cavity and $\frac{8 \cdot 7}{2} = 28$ states with either two photons in different cavities, two excited atoms, or one photon and one excited atom. Similar enumerations can be made for other lattice sizes $N$ and numbers of excitations $N_{\rm exc}$. The dimension ${\cal D}$ grows very rapidly with $N$ and $N_{\rm exc}$.

In one dimension, there are close analogies between quantum spin-$1/2$ models and fermionic tight-binding models as emphasized by the Jordan-Wigner transformation~\cite{jordan28}.  Indeed, the XX spin-$1/2$ Hamiltonian in the sector of total spin $S_z=-N/2+1$  (one up spin in a background of down spins), and the boson and fermion tight binding models in the one-particle sector, have identical single-particle eigenvalues. These connections persist in higher numbers of excitations, with the only distinction being in the allowed occupations of the single-particle levels.

However, in the JCH model, the fact that photon occupations can exceed one, but the emitters are strictly two-level systems negates these connections by mixing spin-1/2 and bosonic operators. An important consequence is that the eigenvalues of the JCH model in the two-excitation sector are {\it not} simply combinations of the single excitation eigenvalues. In this way, the JCH model, despite appearing ``quadratic" in its operators, poses a non-trivial many-body problem in the same way as the periodic Anderson model below, where {\it explicit} quartic interaction terms are present. A more detailed discussion of the many-particle eigenvalue structure of bosonic, fermions, and ``mixed'' operator Hamiltonians is presented in Appendix~\ref{app:eigen}.

The central structure of the JCH model, the coupling of one ``mobile'' collection of quantum degrees of freedom to a set of localized modes is reminiscent of the Kondo ~\cite{anders11} and periodic Anderson models~\cite{sinjukow02,vekic95} used to describe conduction electrons interacting with local magnetic orbitals. The Hamiltonian of the PAM is

\begin{align*}\label{eq:PAMH}
    \hat {\cal H}_{\textnormal{PAM}} &= 
    -\sum_{i,\sigma}
    t_{i}^{\phantom{\dagger}} \big( \, \hat c_{i+1,\sigma}^{\dagger} \hat c_{i\sigma}^{\phantom{\dagger}} + 
     \hat c_{i\sigma}^{\dagger} \hat c_{i+1,\sigma}^{\phantom{\dagger}} \, \big) + \numberthis \\
      \sum_{i,\sigma}
      v_{i}^{\phantom{\dagger}} \big( \, & \hat d_{i\sigma}^{\dagger} \hat c_{i\sigma}^{\phantom{\dagger}}
       +  \hat c_{i\sigma}^{\dagger} \hat d_{i\sigma}^{\phantom{\dagger}} \, \big) 
     +\sum\limits_{i=1}^{N}  U^{c} \, \hat n^{c}_{i\uparrow}\hat n^{c}_{i\downarrow}
     + U^{d} \, \hat n^{d}_{i\uparrow}\hat n^{d}_{i\downarrow} \ .
     \nonumber
\end{align*}

Similar to the JCH model, the length of the system, $N$, is the number of conduction sites. $\hat c_{i\sigma}^{\dagger} \big(\hat c_{i\sigma}^{\phantom{\dagger}} \big)$ are conduction electron creation (annihilation) operators at site $i$ and spin $\sigma$, $\hat d_{i\sigma}^{\dagger}\big(\hat d_{i\sigma}^{\phantom{\dagger}} \big)$ are creation (annihilation) operators of localized electrons at site $i$ and spin $\sigma$, and $\hat n^{c}_{i\uparrow}, \hat n^{c}_{i\downarrow}, \hat n^{d}_{i\uparrow}, \hat n^{d}_{i\downarrow}$ are the associated number operators.

The model is parameterized by conduction electron hopping $t_{i}$, inter-orbital hybridizations $v_{i}$, and on-site repulsions $U^c$ and $U^d$. The geometry of the PAM is illustrated in Fig.~\ref{fig:JCHGeom} (bottom). The structure in the top and bottom panels emphasizes the geometric similarities between the JCH model and the PAM but obscures the fact that the PAM has two spin species of fermionic conduction electrons, whereas the JCH model has a single species of mobile bosons. Similarly, the localized orbital accommodates spin-up and spin-down electrons, unlike the single excitation of the atoms in the JCH. These differences become manifest when $N_{\rm exc}>1$.

The brute force methodology for solving $\hat {\cal H}_{\rm PAM}$ is identical to that for $\hat {\cal H}_{\rm JCH}$: choose a basis, construct the matrix, and diagonalize. The occupation number states for the PAM are, of course, different from those for the JCH, due to the difference in the statistics of the operators and the presence of an additional spin degree of freedom in the PAM.

In the absence of the interaction terms, QST in the PAM with multiple excitations is the same as for a single excitation -- a consequence of the fact that the many-particle eigenenergies are simple sums of the single particle excitations when $U^c=U^d=0$ and all the degrees of freedom share fermionic statistics. As noted above, this is not the case for the mixed character of the degrees of freedom in the JCH model. The effect of interactions on multi-excitation QST in the Hubbard model has been considered in \cite{PhysRevB.73.195122}.


\subsection{Methods}

A high-level description of Monte Carlo methods emphasizes their flexibility:  given a collection of degrees of freedom $\{ \, \phi_i \, \}$ and a probability distribution ${\cal P}\big(\{ \, \phi_i \, \} \big)$ one defines an appropriate transition matrix ${\cal T}\big( \{ \, \phi_i \,\} \rightarrow \{ \, \phi_i \,\}^\prime \big)$ so that, as ${\cal T}$ is repeatedly invoked to generate a sequence of configurations, the degrees of freedom are generated with the desired probability 
    \bibnote{This is done by requiring ${\cal T}$ to obey ``detailed balance'' via, for example, the heat bath or Metropolis algorithms.  This ensures that ${\cal P}\big(\{ \, \phi_i \,\}\big)$ is an eigenvector of ${\cal T}$ of largest eigenvalue. It is not necessary to normalize ${\cal P}$, since only the ratio ${\cal P}\big( \{ \, \phi_i \,\} \, \big/ \, {\cal P} \big( \{ \, \phi_i \,\}^\prime \big)$ is needed. }.

To accomplish efficient QST, our procedure is then as follows. We define a target time evolution operator $\hat {\cal U}^{\, \bf t}$ which, having chosen a basis for our Hilbert space, has matrix elements ${\cal U}^{\, \bf t}_{ab}$ with $a,b$ labeling the basis vectors (rows and columns of $\hat {\cal U}$). $a$ and $b$ run from $1$ to the dimension ${\cal D}$ of the Hilbert space. This time evolution operator is a function of the parameters of the Hamiltonian and of the time, and encodes the desired perfect QST. If we want to go from a specific basis state $|\alpha\rangle$ (e.g.~one which represents excitations initially localized on particular sites) to another specific state $|\beta\rangle$ with excitations at some final locations) we require column $\alpha$ of the matrix for $\hat {\cal U}$ to consist entirely of zeros except for row $\beta$. 

In order to have our Monte Carlo enforce these matrix elements we define an ``action''
that quantifies the difference between the matrix of our current time evolution operator and the target:

\begin{align}
    {\cal S} \equiv \frac{1}{T} \sum_{a,b}^{}{\vphantom{\sum}}^{\prime}
    \, \Big({\, \cal U}^{\, \bf t}_{ab} - {\cal U}_{ab} \, \Big)^{2} .
    \label{eq:S}
\end{align}

Here the prime on the sum emphasizes that it runs only over those matrix elements (columns of ${\cal U}$) which are targeted.  If we want to specify perfect QST for only one pair of initial and final states, the sum in Eq.~\ref{eq:S} is only over one column.  Adding requirements for perfect QST between additional pairs expands the number of terms in the sum for ${\cal S}$. This will be discussed in Sec.~\ref{sec:Multiple Transfers}.

The Monte Carlo then proceeds by making changes to each of the parameters in the Hamiltonian in succession, and accepting those changes with probability given by the ``heat bath'' prescription:
$p = e^{-\Delta {\cal S}}\, / \, \big( 1 + e^{-\Delta {\cal S}} \big)$. Here $\Delta {\cal S}$ is the change in the action induced when the Hamiltonian parameter is altered. Equation \ref{eq:S} contains a ``temperature'' $T$. To achieve perfect QST, one desires to converge to ${\cal S}=0$ so that the time evolution operator exactly matches the target. This would eventually occur if we set $T=0$ to suppress all fluctuations about the target. The purpose of making $T$ finite is to prevent getting caught in metastable states in the course of optimizing the Hamiltonian to perfect QST, and also to allow large Monte Carlo moves to be accepted early in the simulation. An effective annealing schedule~\cite{kirkpatrick83} is to start at a high temperature $T_{\rm max}$ and then reduce $T$ logarithmically in steps towards a small but finite $T_{\rm min}$. A procedure which is found empirically to work reasonably well is 
to start at the hoppings $J_i = 1$ and reduce the temperature from $T_{\rm max} \sim 100 $ to $T_{\rm min} \sim 0.1$.

However, this primitive annealing can be improved. The ``dual annealing'' method, also called ``generalized simulated annealing,'' \cite{RJ-2013-002,pedamallu08} provides better convergence (a factor of 2-3) \cite{JSSv060i06} and hence is used in the results which follow. The ``dual annealing'' method uses the same main accepting and rejecting protocol as the primitive Monte Carlo but includes several variations and additional procedures, as described below. We have implemented some of these changes into our Monte Carlo algorithms to speed up convergence.

The most potent change is that instead of using a uniform distribution to generate random perturbations, it uses the Cauchy-Lorentz distribution, which was shown to speed up convergence in \cite{TSALLIS1996395}. Another key modification is that while the temperature globally decreases over many iterations, it is occasionally ``re-annealed,'' sending the temperature higher for a brief period and increasing the rate of temperature decline. This puts the optimization algorithm into alternating periods of frequent, larger changes, and infrequent, smaller changes. The two frequencies of the temperature annealing process is where the method ``dual annealing'' gets its namesake. An important caveat to this annealing schedule is that the best position is saved before each re-annealing of the temperature so that if no better solution is found during the high-activity period, then the program can return to this last solution and avoid getting sent off track.

We conclude this section by commenting on the computational cost and the explored phase space. Regarding cost, we must re-compute the matrix for $\hat {\cal U}$ after each change in a Hamiltonian parameter. This involves diagonalizing and exponentiating the Hamiltonian, operations that scale with the cube of the Hilbert space dimension ${\cal D}$. As noted earlier, ${\cal D}$ is a rapidly increasing function of $N$ and $N_{\rm exc}$. For the JCH model with $N=8$ and $N_{\rm exc}=2$ we have ${\cal D}=128$, so a sweep through all 16 parameters $\{ \, J_i, g_i \, \}$ lattice takes ${\cal O}\big(16 \cdot (128)^3 \big)\sim 10^7$ operations. Doing $10^4$ sweeps would then represent ${\cal O}(10^{11})$ operations and take tens of minutes on a several GHz processor workstation. Thus, the time grows rapidly with $N$ and $N_{\rm exc}$. However, the system sizes accessible to the theory are still of the same order as those being explored experimentally at present~\cite{Li2022}.

Regarding the phase space, the JCH and PAM have $2N-1$ and $2N+1$ parameters, respectively, which are adjusted to optimize for high-fidelity QST. To transfer each initial and final state pair, there are ${\cal D}$ constraints involved in matching ${\cal U}$ to ${\cal U}^{\, \bf t}$. We can see that as $N$ and $N_{\rm exc}$ grow, the number of constraints begins vastly to exceed the number of adjustable parameters. Thus there is no {\it a priori} guarantee that a good solution will be found. One of our chief messages is that, despite having (many) more equations than unknowns, achieving good QST is still feasible.

\begin{figure}[t] 
\includegraphics[width=1\columnwidth]{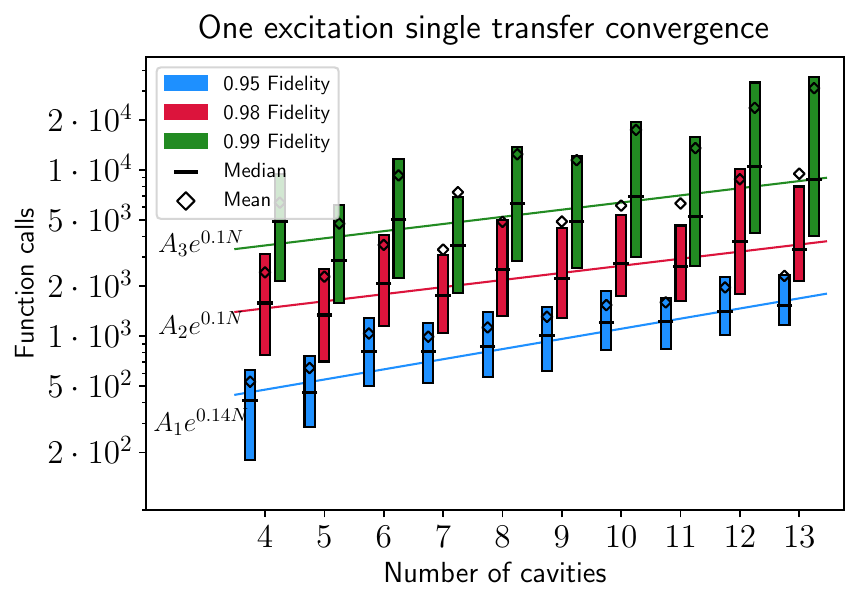}
\caption{
    The distribution of function calls used to reach target levels of fidelity for one excitation systems of various sizes. For each number of cavities, our Monte Carlo methods were run $200$ times with different seeds and the numbers of function calls used at the benchmarks of $0.95$ (blue), $0.98$ (red), and $0.99$ (green) fidelity were recorded. In the plot, each box represents the range between the $25$th percentile and the $75$th percentile of function calls. The black bar marks the median value and the black diamond marks the mean. Note that the $y$ axis is on a log scale. The scaling to reach a fixed fidelity is exponential in $N$. Each line is fit to $A_fe^{ m_f N}$. Here $A_1=272, A_2=994, A_3=2368$. 
}
\label{fig:CCAConvergence}
\end{figure}


\section{QST in the Single-Excitation Sector}\label{sec:Single-Excitation}

Here we will demonstrate our methods on a problem with known solutions -- QST in the single-excitation sector of a 1D coupled cavity array. In this case, perfect QST can be found reliably by utilizing Monte Carlo with an action that targets ``special eigenvalues'' found in \cite{PhysRevB.105.195429}. However, in the next section, we will see that optimizing the time-evolution matrix is a much more versatile method.

When working in these systems, we define our basis to include the states with a single excitation, either a photon in one of the cavities or an excitation of one of the emitters. We define our Monte Carlo's error as the value in our time-evolution matrix representing the probability that a particle will transfer between the leftmost cavity and the rightmost cavity at a time $t_p$.

We observe that these Monte Carlo methods can reliably reach any high fidelity between these two states given enough Monte Carlo iterations. To benchmark our methods, we ran independent optimizations many times with different random seeds and recorded the number of function calls (i.e., the attempt to update a single Hamiltonian parameter, recomputing the corresponding unitary ${\cal U}$) required to converge to various fidelity levels for transfer between the ends of the top row of sites [see Fig.~\ref{fig:JCHGeom}(a)]. These results are displayed in Figure \ref{fig:CCAConvergence}.

We observe that the number of necessary function calls increases as we demand higher fidelity or increase the system's complexity (distance of propagation). The correlation between the number of cavities and function calls appears roughly linear on the log scale, implying an exponential relationship. There also appears to be a small parity bias favoring systems with an odd number of cavities. As is typical with stochastic optimization methods, the variance is quite large. There is no {\it guarantee} that a target fidelity will be reached under any number of function calls.  Note that the computational difficulty in solving larger systems sizes increases both because the optimization requires more function calls, but also because each function call becomes more expensive, as adding more states creates a larger matrix that must be diagonalized to obtain ${\cal U}$. 

\begin{figure}[t] 
\includegraphics[width=1\columnwidth]{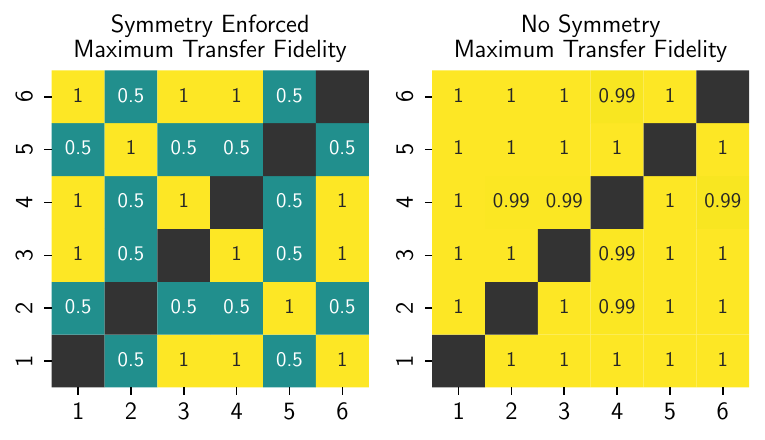}
\caption{
    An $N=3$ coupled cavity array with a single excitation. The six possible states are arranged along the $x$ and $y$ axes. Each square represents a transfer between two states. The value of each square is the maximum fidelity achieved after $3000$ function calls when optimizing for that transfer specifically. \underbar{Left:} Each optimization has coupling where symmetry is enforced. \underbar{Right:} A repeat of this process without enforcing symmetry on the couplings. 
}
\label{fig:CCA3Combined}
\end{figure}

Our method is not limited to QST between the top leftmost and rightmost sites. We can target transfer between any pair of states by specifying a different target time-evolution operator. Aside from transfers between states that violate a conservation or symmetry law, there are many pairings of states to choose from. In Fig.~\ref{fig:CCA3Combined}, we exhaustively demonstrate that all single target transfers can be achieved in a length $N=3$ system with ${\cal D}=6$ total states.

From the left side of Fig.~\ref{fig:CCA3Combined} we observe that the optimization for each state transfer converged to either $0.5$ or to $1$. We note that every value on the anti-diagonal has converged to $1$, meaning we can transfer any state to its ``mirror'' state with perfect fidelity. Some other elements can only reach $0.5$ due to our symmetry constraints. On the right side without such constraints we see all optimizations reach almost $1$. As we primarily focus on symmetric transfers, we will typically apply symmetry constraints to halve the number of optimization parameters.


\section{Multiple Excitations}\label{sec:Multiple Excitations}

We now move on to the main focus of the paper: QST with multiple, interacting excitations. This is a difficult task -- when adding a second excitation in either the JCH or PAM models the basis size is roughly squared, but the number of tunable hoppings stays the same. Thus it is no longer feasible to require perfect state transfer for all system states. Instead we must focus on a smaller number of state pairings and refined methods.


\subsection{Jaynes-Cummings-Hubbard Model}

In the JCH model, the obvious first guess to finding high fidelity QST in the multiple excitation sector is to use solutions that worked with one excitation. However, as shown in Ref.~\cite{PhysRevB.105.195429}, the behavior of a two-excitation system bears little resemblance to that of a one-excitation system with the same coupling values and therefore the one-excitation solution cannot be used. 

Another approach that comes to mind is to apply the eigenvalue targeted optimization method that we used to find perfect QST in the one-excitation JCH model in \cite{PhysRevB.105.195429}. But this requires foreknowledge of the eigenvalues that lead to perfect QST, which we do not have in the two-excitation system. In addition, the eigenvalue patterns in the one excitation system cannot be reached in the two-excitation case, due to the many differences in their respective formats discussed in Appendix~\ref{app:eigen}.

We therefore turn to our Monte Carlo methods that target the time-evolution operator. In the JCH system we will transfer from the state with two (photonic) excitations in the top leftmost cavity to the state with two excitations in the top rightmost cavity to serve as an example of our method. Just as described in Sec.~\ref{sec:Single-Excitation}, we compute the time evolution matrix at time $t_p$ and optimize for our desired values.

\begin{figure}[t] 
\includegraphics[width=1\columnwidth]{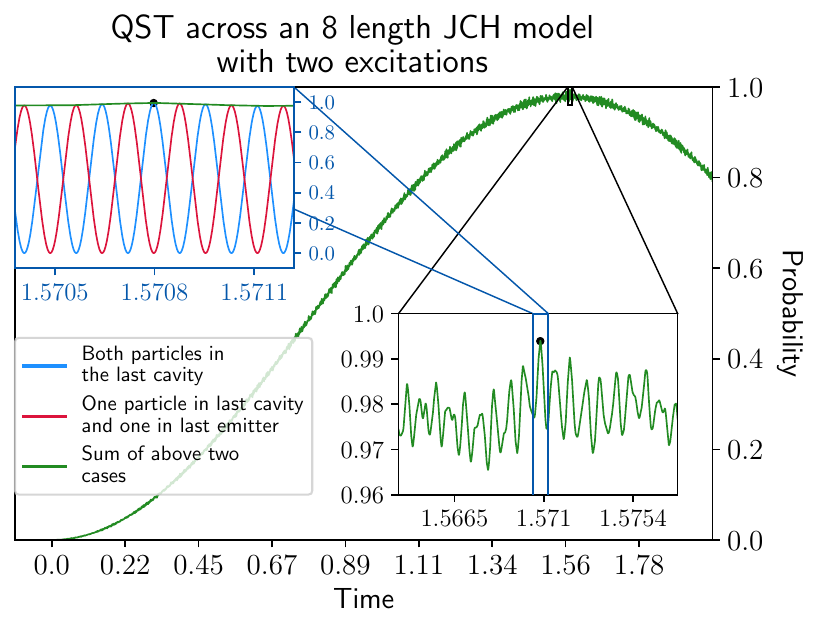}

\textbf{Coupling Values}  \\   
\centering
\begin{tabular}{|c|c|c|c|}
    \hline
    $J_1, J_7$ & $J_2, J_6$ & $J_3, J_5$ & $J_4$ \\
    \hline
    926.29   & 5124.13   & 4273.83 & 9241.97  \\
    \hline
    $g_1, g_8$ & $g_2, g_7$ & $g_3, g_6$ & $g_4, g_5$ \\
    \hline
    15384.66   & 16447.19   & 16654.13 & 2055.53  \\
    \hline
\end{tabular}

\caption{
    Demonstration of high-fidelity transfer between two states in the JCH model with two excitations in $N=8$ cavities.  We consider transfer beginning with two photons co-located in the leftmost cavity and transferred to the rightmost cavity. In the figure, we plot the fidelity of this transfer over time. The peak fidelity is $0.994$. In the top left inset, note that the system oscillates rapidly between having both particles in one cavity and having one particle in the cavity and one in its associated emitter. The period of these oscillations is approximately $10^5$ times smaller than the principal oscillation's period. To illustrate this principal oscillation, the main plot displays the sum of the probability of these two states. This sum is close to a sine wave, with small disordered oscillations visible in the lower right inset.
 }
\label{fig:JCHH8_T1_Transfer}
\end{figure}

While the two-excitation JCH system is significantly harder to optimize than the one-excitation JCH system due to its increased basis size and complexity, our Monte Carlo methods can still achieve high-fidelity QST given enough iterations. For example, in Fig.~\ref{fig:JCHH8_T1_Transfer}, we highlight a solution in an $N=8$ JCH model that achieves a fidelity of $0.994$.

Note that the transfer time $t_p$ can be engineered to any desired value by scaling the coupling values. As found in Ref.~\cite{christandl04}, and is clear from dimensional analysis, multiplying each coupling value by a constant $J_0$ scales the transfer time by $J_0^{-1}$.

It is interesting to note that the evolution of this system has several differences from its one-excitation and cavity-only counterparts. The fidelity over time graph for cavity-only perfect QST was characterized by a single sine wave \cite{christandl04}, and the single excitation JCH model appeared to be a combination of two similar frequencies. However, the solution in Figure \ref{fig:JCHH8_T1_Transfer} contains an extremely rapid oscillation between cavity and emitter that is superimposed upon the main oscillatory behavior. Physically, this means that getting both excitations across the chain is almost guaranteed, but determining whether the emitter is excited requires extremely high precision. In future research, it may be beneficial to create a method that finds solutions without rapid oscillations.

The transfer time is also significantly larger than in the cavity-only and single-excitation systems relative to the scale of the coupling values. In both of those systems, the transfer time was inversely proportional to the scale of all the hopping rates. As some of the coupling rates in the two-excitation system are significantly larger than in the cavity-only and single-excitation solutions, one might expect the transfer time to be shorter, but instead, it is nearly two orders of magnitude larger. Since the two-excitation system has a significantly larger basis size, the longer transfer time may be because more frequencies must coincide simultaneously to get perfect QST. However, the higher coupling values explain the rapid oscillations between the end cavity and emitter.

\begin{figure}[t] 
\includegraphics[width=1\columnwidth]{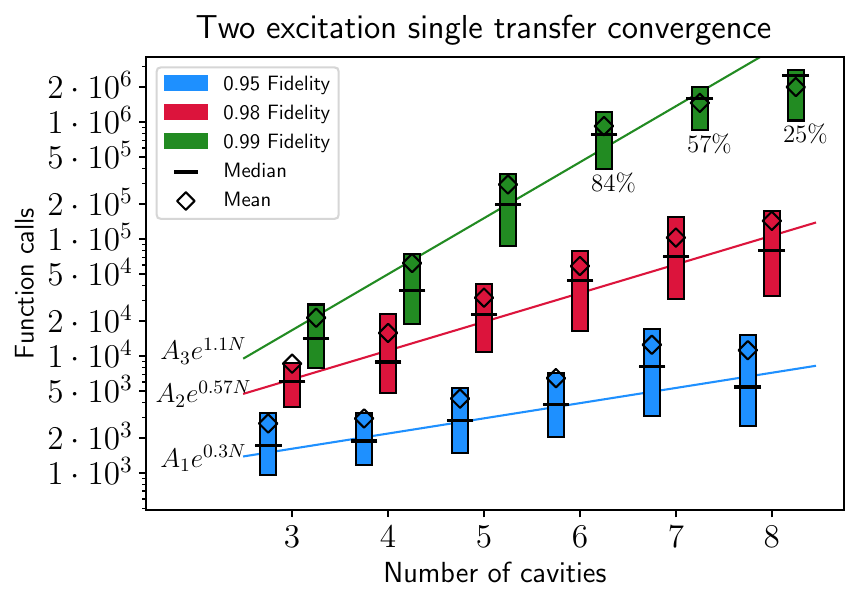}
\caption{
    Distribution of function calls needed to reach target levels of fidelity for two-excitation systems. In each system, we target the transfer of two particles localized in the top left cavity transferring to the top right cavity. 
    Our Monte Carlo methods were tested in the same manner as in Fig.~\ref{fig:CCAConvergence}.
    Each optimization run was given a maximum of $400,000\cdot(\text{number of independent variables})$ function calls before termination. The percentage of runs that reached the benchmark within the maximum function calls is annotated below each bar when it is not $100\%$ Note that the $y$ axis is on a log scale. Each line is fit to $A_fe^{ m_f N}$. Here  $A_1=657, A_2=1161, A_3=610$. 
}
\label{fig:JCHH_2_Convergence}
\end{figure}

We can benchmark the number of function calls required to reach high-fidelity transfer in the two-excitation case as well. This is recorded in Fig.~\ref{fig:JCHH_2_Convergence} using the same methods as the convergence plot for the one-excitation case in Fig.~\ref{fig:CCAConvergence}. While the exponential dependence on the number of cavities persists, the prefactor $m_f$ in the exponential is typically much larger, rendering much more costly annealing. This indicates that the landscape of solutions that minimize ${\cal S}$ is shallower and that the number of solutions that reach the target fidelity is reduced -- this is characteristic of increasingly quantum chaotic systems.

\subsection{Periodic Anderson Model}

We now move on to the Periodic Anderson Model. In this model, adding a second excitation (of opposite spin) will introduce correlations due to the interaction constants $U^c$, $U^d$ in the Hamiltonian. Without such interactions, the system is identical to two separate one-excitation systems, and the solution is unchanged. 

In Ref.~\cite{PhysRevB.73.195122}, it was found that in the single-band Hubbard model, near-perfect QST could be achieved using the single excitation solution and certain optimal values of the interaction $U$. This prompted us to search for similar values in the PAM. We begin with the solved coupling values for the single-excitation system and manually search for special $U^c, U^d$ values that exhibit high-fidelity QST with two excitations. We applied this method for an $N=3$ length system, but we see that even for such a small case, we do not find nontrivial $U^c, U^d$ that allow a one-excitation solution to work with two interacting excitations. In a sweep of both $U^c$ and $U^d$ from $0$ to three times the average coupling value, the highest nontrivial solution had fidelity $0.229$ when transferring one up and one down electron together across the top chain and $0.817$ when swapping the positions of a top and down electron placed at opposite ends of the top chain. Thus, it is clear that we must find entirely new solutions.

\begin{figure}[t] 
\includegraphics[width=1\columnwidth]{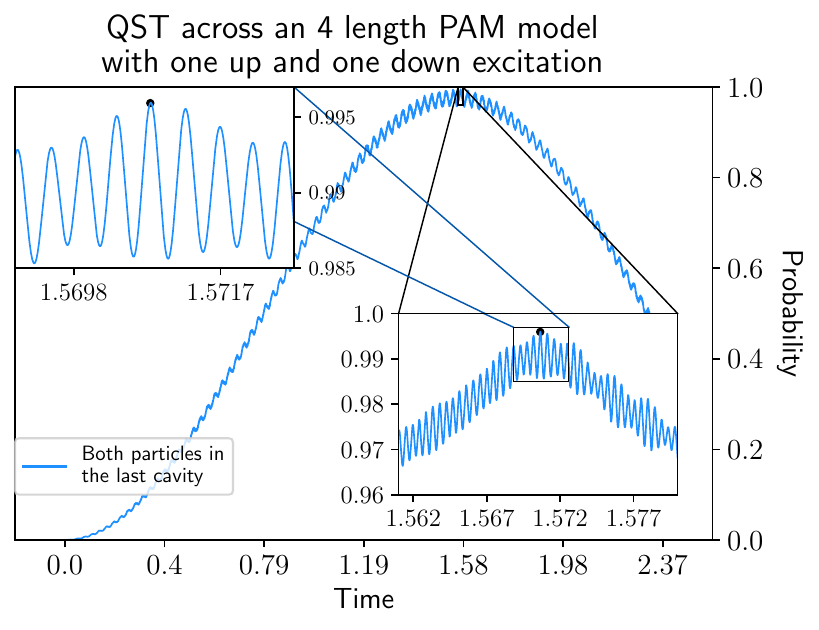}

\textbf{Coupling and interaction values}  \\
\centering
\begin{tabular}{|c|c|c|c|c|c|}
    \hline
    $t_1, t_3$ & $t_2$ & $v_1, v_4$ & $v_2, v_3$ &  $U^c$  & $U^d$ \\
    \hline
    419.87   & 1188.11  & 478.51 & 1292.36 & 13683.6   & 4014.55  \\
    \hline
\end{tabular}

\caption{
    High-fidelity transfer between two states in the $N=4$ PAM with one spin-up fermion and one spin-down fermion. We used our Monte Carlo algorithms to target transfer between the state corresponding to both fermions localized in the top left site and the state corresponding to both in the top right site. This transfer is challenging as the on-site repulsion components of the Hamiltonian tend to discourage the particles from localizing in the same site. It is important to note that although the optimization algorithm can tune these repulsion constants, the final result does not minimize these values. The system reaches a peak fidelity of $0.996$. 
}
\label{fig:PAM4_T1_Transfer}
\end{figure}

We move on to applying our time-evolution matrix targeting Monte Carlo methods to the PAM systems. These methods are implemented in a similar way as the JCH system; we vary our coupling values, create and diagonalize the Hamiltonian, and compute the error from the time evolution matrix. We find that our prescription can still find high fidelity QST in the PAM system, as demonstrated in Fig.~\ref{fig:PAM4_T1_Transfer}. In this solution, we achieve fidelity of $0.996$. 

The behavior of this system over time appears to follow a sinusoidal path with an additional oscillation of higher frequency. In contrast to the JCH model, we see much lower amplitude rapid oscillations in the transfer fidelity. This oscillation is between the target state with both fermions in the top right site and a equal superposition of the two states with one fermion in the top right site and one in the rightmost offshoot site (conduction and localized orbitals). A benefit to this is that any measurements will be less impacted by imprecision at the measurement time. The transfer time is of a similar order of magnitude to the two-excitation JCH solution. 

One difficulty we encounter when searching for solutions in the PAM system is that we must avoid the ``trivial solution.'' When both $U^c=U^d=0$ the system is effectively interactionless and the one-excitation solution reemerges. To counteract this, we placed bounds on how small the interaction coupling values can be. In our algorithms, we discouraged the trivial solution by increasing the calculated error of proposed coupling values if either of the on-site interactions were less than $0.1$ times the average strength of the coupling values.

In terms of convergence, although our optimization method is the same as for the JCH model, between a PAM system and a JCH system of equal length, the PAM system will be more computationally expensive to optimize, assuming the two excitations have opposite spin.  In that case, the excitations in the PAM model are distinguishable, and the basis will be roughly twice as large, meaning eight times more difficult to diagonalize. In addition, the PAM has two more independent variables that must be optimized, namely the interaction constants $U^c$, $U^d$.

Ultimately we conclude that we can find high-fidelity solutions for the PAM with multiple interacting particles. We found high-fidelity transfers for multiple pairs of states in PAM systems with two electrons and up to $10$ total sites. This further shows the efficacy of this method in optimizing QST in a large variety of challenging models. 

\section{Multiple Transfers}\label{sec:Multiple Transfers}

In this section, we will target Hamiltonians which achieve good QST between several different pairs of initial and final states, with multiple excitations. As we have detailed in our methods section, we can select any number of matrix elements to be optimized in our Monte Carlo. However, the difficulty of finding a solution increases with the number of constraints. We reliably found high-fidelity QST when targeting only one pair of states in Sec.~\ref{sec:Multiple Excitations}, but on the other extreme, our searches indicate that there is no good solution whereby every state is reflected across the center (although such a ``complete'' solution is possible in the single-excitation sector \cite{PhysRevB.105.195429}). We will now investigate how many sets of states we can transfer with one single set of coupling values. 

\begin{figure}[t] 
\includegraphics[width=1\columnwidth]{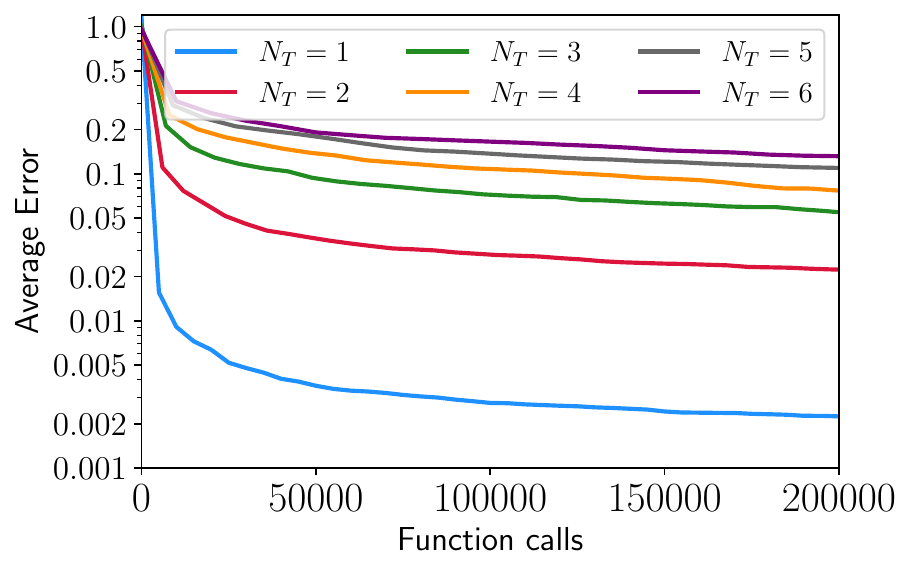}
\caption{
Average convergence of fidelity for different numbers of target transfers $({N_T})$ in an $N=4$ JCH system. The $x$ axis is the number of function calls. The average error along the $y$ axis is one minus the average transfer fidelity, averaged over one hundred independent optimizations. Note that these average errors are far above the lowest errors we can find given enough iterations. Indeed, the best average fidelities of our $100$ runs for $N_T$ transfers one through six were $(1.000, 0.994, 0.991, 0.986, 0.976, 0.972)$, respectively. So while it is the case that an optimization with fewer target transfers usually converges more rapidly, we emphasize that we can eventually achieve high fidelity in the systems with many target transfers.
}
\label{fig:multipleTransferConvergence}
\end{figure}

To test the impact of adding constraints to our Monte Carlo algorithms, we run a series of optimizations on an $N=4$ JCH system with $N_{\rm exc} = 2$. First, we create a list of all desired pairs of states to transfer between by pairing all states with their mirror image states (ignoring self-symmetric ones). Then, we run many independent optimizations for different numbers of target transfers. On each run with $N_T$ target transfers, we randomly selected $N_T$ pairs of states from our set. When calculating the final error we divide by $N_T$ so that the error is one minus the average fidelity of all transfers. The results of these optimizations are shown in Fig.~\ref{fig:multipleTransferConvergence}.

As expected, adding more target transfers (more constraints), utilizing the same set of couplings, results in lower average fidelity. The decrease from $N_T=1$ to $N_T=2$ is the most dramatic, meaning we cannot guarantee very high fidelity for anything more than one target transfer. However, even with six transfers, we can achieve an average fidelity of eighty percent after just $2\cdot 10^6$ function calls, which is promising. In real-world applications, the fidelity may be higher than seen here--we have chosen random transfers for this assessment, but the fidelity is typically much higher when the target transfers are similar to each other. 

\section{Conclusions}\label{sec:Conclusions}

We have demonstrated the efficacy of a novel Monte Carlo/dual annealing method at finding high-fidelity QST in systems with multiple, interacting excitations. To optimize the fidelity between two states in a system, we used tailored Monte Carlo algorithms that optimize our coupling values to produce the time-evolution matrix corresponding to our transfer. In addition, we have benchmarked the convergence in the JCH model with and without multiple excitations to quantify the difficulty that interacting excitations have on engineering good QST. Lastly, we gauged the limits of our optimizations by solving for increasing numbers of state transfers in the same system.

We note that one can also perform a Monte Carlo in the Hamiltonian parameters to minimize the ``infidelity'', $1 - |\langle \psi(t_{\rm QST} | \psi_{\rm target} \rangle |^2 $, that is, the difference between the wave function at some QST transfer time and a target  wavefunction \cite{QST_our_exp}. This latter approach has some potential advantages in scaling with the Hilbert space dimension, since one can use Lanczos-type algorithms to compute the action of ${\cal U}$ on the initial wave function rather than constructing the full matrix of ${\cal U}$. Exploration of the most efficient optimization procedure is an interesting area for future work.

Initial time evolution studies in the Jaynes-Cummings-Hubbard model in the single-excitation sector are described in \cite{PhysRevB.105.195429}. There, a different approach to determining the Hamiltonian was used, one in which minimization of the action altered the hoppings such that a set of target eigenvalues was achieved. The method proposed here is much more powerful for two reasons.  Most importantly, while the eigenvalues that will result in good QST are known in certain special cases \cite{christandl04}, they are not known generally. The approach described here eschews the need for preknowledge of the eigenvalues and focuses only on the actual objective, namely QST. It also has the flexibility to search for different types of evolution (multiple pairs of initial and final states). Meanwhile, the practicality of the two methods is comparable. Both involve multiple diagonalizations of a ${\cal D}$ dimensional matrix, and both involve tuning ${\cal O}(N)$ free parameters in order to achieve ${\cal O}({\cal D})$ constraints (a column of $\hat {\cal U}$ in the approach described here, the eigenvalues of $\hat {\cal H}$ in the approach of Ref.~\cite{PhysRevB.105.195429}). 

In recent years, high-fidelity state transfer has been a central objective of the quantum information science community, explored experimentally in coupled atoms and photons~\cite{matsukevich2004quantum, PhysRevB.86.195312} using microwave photons~\cite{kurpiers2018deterministic}, quantum dots~\cite{Kandel2021} and with superconducting qubits~\cite{bienfait2019phonon, QST_our_exp}.

In particular, experimental realizations of the JCH model have progressed in various platforms~\cite{PhysRevLett.111.160501, Li2022}. As quantum computers increase in scale, the high-fidelity transfer of states will be critical to achieving an interconnected quantum network. Our Monte Carlo approach provides a path for optimizing couplings, which is practical for networks of the same size as those currently constructed. The work reported here was performed on a small collection of standard workstations, allowing the possibility of substantially larger systems using more powerful hardware, even though the computational cost scales exponentially.

A reasonable extension that follows from the versatility of the methods we outlined in this paper is to apply these methods to other models, for example, the Kondo and Hubbard Hamiltonians. Another interesting extension is the application of the methods developed here to higher dimensions. Initial work with 2D superconducting qubit arrays has been undertaken in Ref.~\cite{QST_our_exp}. The results reported there point to the generality of using Monte Carlo to engineer high-fidelity QST between desired states in higher dimension.


\vskip0.20in \noindent
\underbar{Acknowledgements}
A.Y.~and R.T.S.~were supported by the grant DE‐SC0014671 funded by the U.S. Department of Energy, Office of Science. 
R.M.~acknowledges support from the NSFC Grants No.~NSAF-U2230402, 12111530010, 12222401, and No.~11974039.
A.Y.~conducted the simulations.  All authors participated in the ideas behind the research, the analysis of the data, and the writing of the manuscript.


\newpage
\appendix

\section{Eigenvalues of bosonic, fermionic, and mixed  statistics}
\label{app:eigen}

Prior investigations of Monte Carlo optimization of QST \cite{PhysRevB.105.195429} were centered around the eigenvalues of the Hamiltonian. This numerical work generalized the discussion of \cite{christandl04}, which found that integer spacing between the ratios of successive eigenvalues in a one-excitation system's Hamiltonian was both a necessary and sufficient condition for that system to exhibit perfect QST.

While we have emphasized in this paper that targeting the time-evolution operator $\hat {\cal U}$ via Monte Carlo appears to be the more flexible and general method for achieving high-fidelity QST, in this Appendix, for additional insight, we explore two features of the eigenvalue distribution for two-excitation QST:  (i) a comparison of the ratios of the eigenvalues after the Monte Carlo has optimized ${\cal U}$ to initial random couplings; and (ii) the relation between two excitation and single excitation eigenvalues for the JCH model.

\noindent
\subsection{Eigenvalue properties after optimization of $\hat {\cal U}$}

One indication that eigenvalues still remain pertinent to QST in systems with multiple, interacting excitations is an analysis of the ratios between successive eigenvalues for optimized Hamiltonians and for random ones. As found by \cite{christandl04}, the ratios of spaces between eigenvalues is critical for QST in the one-excitation sector. We examine a large sample of independently optimized coupling values and determine if the ratios between successive eigenvalues follow a different distribution than when the coupling values are chosen randomly. Indeed, we noticed a distinct difference between the two distributions—this ratio is close to $1$ significantly more often in systems of optimized couplings than in systems of random couplings. A deeper analysis of systems optimized for QST using random matrix theory is done in \cite{QST_our_exp}.

\begin{figure}[t] 
\includegraphics[width=1\columnwidth]{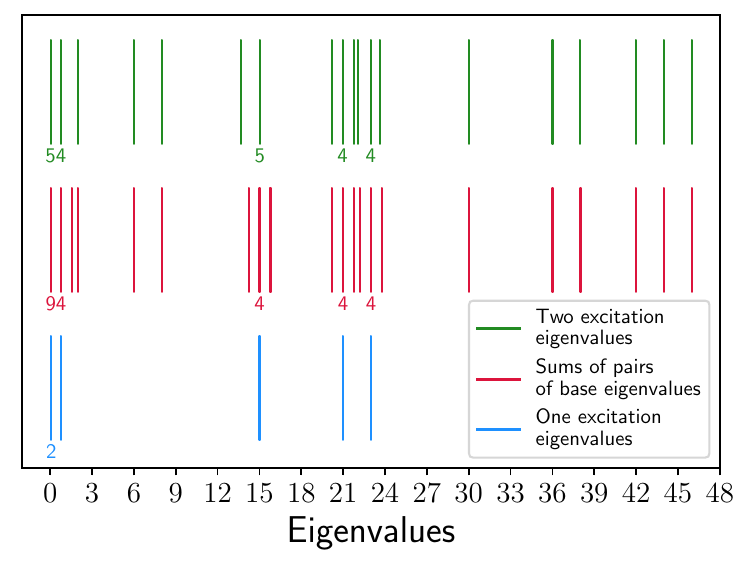}
\textbf{Coupling Values}  \\   

\centering
    \begin{tabular}{|c|c|c|c|c|c|}
    \hline
    $J_1,J_5$ & $J_2,J_4$ & $J_3$ \\
    \hline
    20.531   & 5.309   & 16.228   \\
    \hline
    $g_1,g_6$ & $g_2,6_5$ & $g_3,6_4$  \\
    \hline
    1.023   & 1.544   & 3.645   \\
    \hline
\end{tabular}

\caption{
    This figure highlights the pattern of eigenvalues in our six-length JCH model solved for many transfers. In blue, we plot the $12$ eigenvalues for our optimized Hamiltonian with one excitation. Only the nonnegative eigenvalues are displayed in the plot, but each eigenvalue comes with a negative counterpart. Degenerate eigenvalues are plotted together and labelled with their multiplicity. In red, we plot the $78$ linear combinations of two of the $12$ base eigenvalues. In green, we plot the $72$ eigenvalues of the system with two excitations. Note that the green two-excitation eigenvalues overlap extremely closely with the red combinations of the base eigenvalues. This pattern is not typically observed for systems that do not exhibit QST.
}
\label{fig:JCHHEigenvalueOverlap}
\end{figure}

Another interesting pattern emerges in certain small systems. In our optimizations for these systems we find some some rare solutions with many high fidelity transfers. For instance, in a $6$ length system we found a system that has $9$ unique simultaneous high fidelity transfers (those being the states where excitations in the cavities are transferred to their horizontally mirrored states). In such a system, we would expect that any patterns in the eigenvalues that relate to QST would be more apparent.

\noindent
\subsection{Multi-excitation and single-excitation eigenvalues relationship}

It is well known that Hamiltonians which are quadratic in the creation and destruction operators have a many-particle spectrum that is obtained trivially from the single-particle one, $E^{(1)}_{\alpha}$. If the operators are fermionic, the two-particle eigenvalues are $\{ E^{(2)} \} =  E^{(1)}_{\alpha} + E^{(1)}_{\beta}$, with $\alpha > \beta$, while if the operators are bosonic the same relation holds with $\alpha \geq \beta$.  The fact that $\alpha < \beta$ is not counted separately from $\alpha > \beta$ reflects the indistinguishabilty of the quantum particles, while the Pauli exclusion principle removes $\alpha = \beta$ for fermions. In considering the spectrum of the JCH model, which is quadratic but has operators of mixed character (bosonic photons and fermionic emitters), it is natural to ask if the collection of two excitation eigenvalues $\{ E^{(2)} \}$ nevertheless have any relationship to $E^{(1)}_{\alpha}$. (It is clear the eigenvalues can never be precisely given from single excitation combinations, as the ``blocking'' of excitations in emitters reduces the number of possible states, and hence the number of eigenvalues is different.)

To address this question, we first optimized $\hat {\cal U}$ for high-fidelity QST for an $N=6$ JCH system with {\it two} excitations, and then computed the resulting eigenvalues. Next, we computed the eigenvalues for the same system and the same coupling values but with {\it one} excitation. In this case, we indeed see a high overlap of eigenvalues between the multiple excitation system and combinations of the base system eigenvalues.
    \bibnote{We found that this attribute of high overlap of eigenvalues is {\it descriptive} but not {\it prescriptive} of high fidelity QST as there exist many systems with this quality do not exhibit high fidelity QST.}.

Despite the fact that the Hamiltonian appears to be quadratic, the operators from which it is built do not all share a common set of (bosonic or fermionic) commutation relations. Thus the many particle eigenenergies and eigenfunctions cannot be built in a simple way from the one particle sector, as is the case with a purely bosonic or fermion quadratic Hamiltonian.

While we have presented data here for the JCH model, the application of these eigenvalue analyses on the Periodic Anderson Model is a interesting subject for further study. In the PAM, interactions are introduced through the on-site interactions $U^c, U^d$. This form of interactions has the benefit of not changing the basis size, which keeps the number of eigenvalues unaltered. However, the effect of the interaction constants $U^c, U^d$ on the eigenvalues appears to be significantly more complex than in the JCH model, and no patterns associated with high fidelity QST have been observed during optimization.


\newpage
\bibliography{mybib}

\begin{thebibliography}{39}
\bibitem [{hba()}]{hbar}%
  \BibitemOpen
  \href@noop {} {}\bibinfo {note} {Here and throughout this paper we set
  $\hbar=1$.}\BibitemShut {Stop}%
\bibitem [{1()}]{1}%
  \BibitemOpen
  \href@noop {} {}\bibinfo {note} {The Christandl prescription~\cite
  {christandl04} automatically accomplishes perfect QST between all pairs
  representing transitions between states located symmetrically about the chain
  center, as might be inferred from the symmetry of the $\{ \, J_i \,
  \}$.}\BibitemShut {Stop}%
\bibitem [{2()}]{2}%
  \BibitemOpen
  \href@noop {} {}\bibinfo {note} {This is done by requiring ${\cal T}$ to obey
  ``detailed balance'' via, for example, the heat bath or Metropolis
  algorithms. This ensures that ${\cal P}\big (\{ \, \phi _i \,\}\big )$ is an
  eigenvector of ${\cal T}$ of largest eigenvalue. It is not necessary to
  normalize ${\cal P}$, since only the ratio ${\cal P}\big ( \{ \, \phi _i \,\}
  \, \big / \, {\cal P} \big ( \{ \, \phi _i \,\}^\prime \big )$ is
  needed.}\BibitemShut {Stop}%
\end{thebibliography}

\end{document}